\documentclass[letterpaper, 10 pt, conference]{ieeeconf}  

\IEEEoverridecommandlockouts                              

\usepackage{graphics}       
\usepackage{epsfig}         
\usepackage{times}          
\usepackage{amsmath}        
\usepackage{amssymb}        
\usepackage{color}
\usepackage{array}
\usepackage{accents}

\usepackage{stackengine}
\usepackage{url}
\usepackage{enumerate}

\title{\LARGE \bf Experimental Benchmarking of \\ Energy-saving
Sub-Optimal Sliding Mode Control}

\author{Michael Ruderman
\thanks{Michael Ruderman is with University of Agder, Department of Engineering
Sciences, 4879 Grimstad, Norway. \newline
Email: {\tt\small michael.ruderman@uia.no}}%
\thanks{\textcolor[rgb]{0.00,0.00,1.00}{AUTHOR'S ACCEPTED MANUSCRIPT, IEEE VSS2024}}
}

\begin{document}

\maketitle
\thispagestyle{empty}
\pagestyle{empty}

\begin{abstract}
The recently introduced \emph{energy-saving} extension of the
\emph{sub-optimal sliding mode control} allows for
\emph{control-off} phases during the convergence to second-order
equilibrium. This way, it enables for a lower energy consumption
compared to the original sub-optimal sliding mode (SM) algorithm,
both commutating a discontinuous control signal. In this paper,
the energy-saving sub-optimal SM control is experimentally
benchmarked against a standard second-order SM controller which
also has a discontinuous control action. Here the so-called
terminal second-order SM algorithm is used. The controlled plant
is affected by the matched bounded disturbances which are unknown,
and the output is additionally subject to the sensor noise.
Moreover, a first-order actuator dynamics can lead to chattering,
which is parasitic for SM applications. For a fair comparison, the
same quadratic terminal surface is designed when benchmarking both
SM controllers. Both experimentally compared SM algorithms have
the same (bounded) control magnitude and states initial
conditions.
\end{abstract}

\bstctlcite{references:BSTcontrol}

\newtheorem{theorem}{Theorem}
\newtheorem{rem}{Remark}

\section{Introduction}
\label{sec:1}

Sliding mode (SM) control techniques are among the most promising
robust control strategies, as they enable the handling of weakly
known and perturbed systems. While SM control, as methodology, was
first introduced to a larger auditorium in \cite{utkin1992},
followed by several well-perceived textbooks, e.g.
\cite{edwards1998,perruquetti2002,shtessel2014}, we also refer to
a more recent text \cite{utkin2020} that summarizes the basic SM
principles and ongoing issues.

In SM methods, the control authority aims to force an uncertain
system dynamics onto a specified manifold $\sigma$, called also
the \emph{sliding manifold}, and then maintain the system behavior
by staying on it. In this way, an SM control provides
reduced-order dynamics and insensitivity to both, the
uncertainties in system parameters and the external perturbations
under certain conditions. Being in the first-order SM means
ensuring $\sigma=0$ for all times $t > T_c$, that means after
reaching the sliding manifold at $0 < T_c < \infty$. For
second-order sliding modes \cite{levant1993}, also the time
derivative of the sliding variable is forced to zero, i.e.
$\sigma=\dot{\sigma}=0$, thus enabling for robust regulation of
the systems also with relative degree two between the sliding
variable and control value. Here it is worth recalling that the
second-order SM controllers are seen to belong to a more general
class of \emph{higher-order sliding mode} control, actively
developed over the last two and a half decades for the uncertain
systems with the relative degree two and higher, see e.g.
\cite{shtessel2014,utkin2020} and references therein. Applications
and experimental case studies with the use of the SM control (and
also estimation) techniques are numerous. Here to mention just a
few of the recent, the SM applications can be found in e.g. the
temperature control of chemical fluids for silicon wafer
production \cite{koch2020sliding}, in the states estimation for
frequency regulation and economic dispatch in the power grids
\cite{rinaldi2021sliding}, in the control of hydro-mechanical
drives \cite{estrada2024super}, and many others.

The so called sub-optimal SM control was proposed in
\cite{bartolini1997} as a second-order SM technique which does not
require $\dot{\sigma}$ for a discontinuous control commutation.
Recall that a single commutation at the well-defined time instant
in an unperturbed system results in the time optimal
\emph{bang-bang control}, cf. \cite{fuller1960relay}. In turn, the
sub-optimal SM control \cite{bartolini1997} converges robustly to
$(\sigma,\dot{\sigma})$ origin in presence of the bounded
perturbations, while executing a commutating control sequence
$u(t)$ of an increasing frequency. For a detailed survey of the
properties, generalizing extensions, and applications of the
sub-optimal SM control we refer to \cite{bartolini2003}.

Since the sub-optimal SM control $u(t)$ is continuously
commutating between two sign-alternating discrete values $\forall
\: t > 0$, actually like any discontinuous SM control, the
corresponding cumulative energy consumption is proportional to the
execution time, or at least to the convergence time if the control
can be switched off after reaching the origin. Recently, an energy
saving extension of the sub-optimal SM control was introduced in
\cite{ruderman2023energy}, which allows for the well defined $u=0$
phases during the convergence. The original sub-optimal SM control
was modified in \cite{ruderman2023energy} by introducing an
additional switching threshold which enables for a mandatory $u=0$
phase between two consecutive extrema of the sliding variable, cf.
with \cite{bartolini2003}. This implied the control value to be in
the finite set $u \in \{-U,\, 0,\, U \}$, while the class of
perturbed systems with relative degree two and discrete bounded
control, i.e. $\max|u| = U$, were targeted.

In this paper, the energy-saving sub-optimal SM control
\cite{ruderman2023energy} is experimentally benchmarked in
comparison with a classical terminal second-order SM control. The
latter assumes $u \in \{-U,\, U \}$ and provides a finite-time
convergence while compensating robustly for the bounded
perturbations $|d(t)| < U$, see e.g. \cite{ruderman2021}. Worth
noting is also that the energy-saving sub-optimal SM control
\cite{ruderman2023energy} is evaluated experimentally for the
first time, while the targeted electro-magneto-mechanical system
\cite{ruderman2022motion} represents a sufficiently challenging
application case with regard to the uncertainties, disturbances,
and sensor noise.

The rest of the paper is as follows. In Section \ref{sec:2}, we
provide the necessary preliminaries of the energy-saving
sub-optimal SM control \cite{ruderman2023energy} and the
benchmarking terminal SM control with the corresponding
second-order sliding surface. Also the chattering effects are
briefly addressed, since being relevant for the real SM and owing
to an additional actuator dynamics, see e.g. \cite{boiko2009}.
Section \ref{sec:3} describes in detail the experimental
benchmarking system. The experimental results are shown in Section
\ref{sec:4} while benchmarking both, convergence of the output
value of interest and the monitored energy consumption. The latter
is measured as an integral of the control absolute value, i.e.
proportional to the overall time when the control is on, i.e. $u
\neq 0$. The brief conclusions are drawn in Section \ref{sec:5}.

\section{Preliminaries}
\label{sec:2}

In this section, after specifying the class of the systems under
consideration, the energy-saving sub-optimal SM control
\cite{ruderman2023energy}, which constitutes an extension of the
original sub-optimal SM control \cite{bartolini1997}, is
summarized for convenience of the reader. Then, the so called
terminal SM control \cite{zhihong1994} is recalled, while the
emphasis is on design of an optimal second-order sliding surface
discussed in \cite{ruderman2021}. By the end, the appearance and
parameters of chattering, due to additional (parasitic) actuator
dynamics, are briefly addressed.

The class of uncertain dynamic systems under consideration is the
one which can be assigned with the well-defined sliding variable
\begin{equation}\label{eq:2:1}
\sigma = \sigma (\mathbf{x}, t)
\end{equation}
that has the relative degree $2$ with respect to the control input
$u \in \mathbb{R}$. The vector of the system states, assumed as
measurable, is $\mathbf{x} \in \mathbb{R}^n$ with $n \in
\mathbb{N}$ and $2 \leq n < \infty$. Then, the dynamics of the
sliding variable, which is affected by the matched bounded
perturbations $d$, can be written as
\begin{equation}\label{eq:2:2}
\ddot{\sigma}(t) = u(t) + d(\cdot,t),
\end{equation}
where $d$ is an unknown exogenous term. The system perturbations
are supposed, however, to satisfy the global boundedness condition
\begin{equation}\label{eq:2:3}
\bigl | d(\cdot,t)  \bigr | \leq D,
\end{equation}
where $D > 0$ is a known real constant value. Another essential
boundary of the system class \eqref{eq:2:1}, \eqref{eq:2:2} is
\begin{equation}\label{eq:2:3}
\max | u | = U > D,
\end{equation}
that is due to a saturated control action.

\subsection{Energy-saving sub-optimal SM control} \label{sec:2:sub:1}

The sub-optimal second-order SM control, cf. \cite{bartolini2003},
\begin{eqnarray}
\label{eq:2:1:1a}
  u(t) &=& -\gamma(t) \, U \, \mathrm{sign}(\sigma - \beta \sigma_M),
  \\[1mm]
\label{eq:2:1:1b}
  \gamma(t) &=& \left\{%
\begin{array}{ll}
    1,        & \hbox{ if } (\sigma - \beta \sigma_M) \sigma_M \geq 0
    \\[1mm]
    \gamma^*, & \hbox{ if } (\sigma - \beta \sigma_M) \sigma_M < 0 , \\
\end{array}%
\right.    \\
\label{eq:2:1:1c}
  \beta & \in & [0;\, 1).
\end{eqnarray}
was initially proposed in \cite{bartolini1997}. Here, $\gamma^*
\geq 1$ and $\beta$ are the modulation and anticipation factors,
correspondingly. The dynamic state $\sigma_M$ constitutes the last
\emph{extreme value} of $\sigma(t)$ during the control
\eqref{eq:2:1:1a} operates on the sliding variable. The control
parameters have to satisfy, cf. \cite{bartolini2003},
\begin{eqnarray}
\label{eq:2:1:2}
  U & > & \frac{D}{K_m},\\[1mm]
  \gamma^* & \in & [1; \infty) \cap \Biggl ( \frac{2D + (1-\beta)K_M U}{(1+\beta)K_m U}; \infty
  \Biggr).
\label{eq:2:1:3}
\end{eqnarray}
Recall that for a generic case of the sub-optimal SM controllers
\cite{bartolini2003}, the perturbed dynamics is given by
\begin{equation}\label{eq:2:1:4}
\ddot{\sigma}(t) = K(\cdot,t) \, u(t) + d(\cdot,t) \; \; \hbox{
with } \;\; 0 < K_m \leq K \leq K_M,
\end{equation}
cf. with \eqref{eq:2:2}, meaning the control input gain can also
be uncertain, within the known boundaries $K_m$ and $K_M$. Also
recall that the condition \eqref{eq:2:1:2} represents the control
authority that is required to overcome the unknown yet bounded
perturbation. The condition \eqref{eq:2:1:3} ensures the
convergence of $\sigma(t)$ to the origin. Here we note that first
when $(\sigma, \dot{\sigma})=\mathbf{0}$, the second-order SM
appears in its proper sense.

One can recognize that for a constrained control action
\eqref{eq:2:3} on \eqref{eq:2:2}, the parametric condition
\eqref{eq:2:1:3} vanishes, leading to
\begin{equation}\label{eq:2:1:5}
\gamma^* = 1.
\end{equation}
The resulted (simplified) sub-optimal SM control
$$
u(t) = - U \mathrm{sign}\bigl(\sigma(t) - \beta_1
\sigma_M(t)\bigr),
$$
with $\beta_1 \equiv \beta$ served as a starting point for
developing the energy-saving sub-optimal SM control
\cite{ruderman2023energy}.

\begin{figure}[!h]
\centering
\includegraphics[width=0.65\columnwidth]{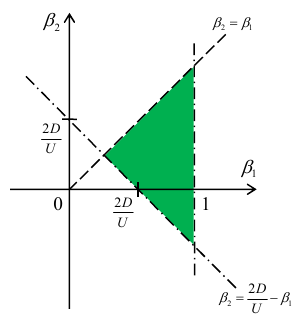}
\caption{Parametric constraints
\eqref{eq:2:1:8}--\eqref{eq:2:1:10} of the energy-saving
sub-optimal SM control \cite{ruderman2023energy} drawn in the
$(\beta_1,\beta_2)$ plane.} \label{fig:21}
\end{figure}
The energy-saving sub-optimal SM algorithm
\cite{ruderman2023energy} is
\begin{equation}\label{eq:2:1:6}
\tilde{u}(t) = -\frac{1}{2} U \mathrm{sign}(\sigma - \beta_1
\sigma_M) - \frac{1}{2} U \mathrm{sign}(\sigma - \beta_2
\sigma_M).
\end{equation}
An initializing control action, similar as in the case of the
original sub-optimal SM control \cite{bartolini2003},
\begin{equation}\label{eq:2:1:7}
\tilde{u}_0(t) = - U \mathrm{sign} \bigl(\sigma(t) - \sigma(0)
\bigr), \quad \forall \, t \in [0;t_{M_1}]
\end{equation}
is also required for accelerating the reaching of the first
extremum at $t_{M_1}$. In addition to the disturbance condition
$|d| < U$, an appropriate thresholds relationship is required,
while the conditions for convergence were derived in
\cite{ruderman2023energy} as
\begin{equation}\label{eq:2:1:8}
\beta_1 + \beta_2 > \frac{2 D}{U},
\end{equation}
\begin{equation}\label{eq:2:1:9}
0  \leq \beta_1 <  1,
\end{equation}
\begin{equation}\label{eq:2:1:10}
-1 < \beta_2 < \beta_1.
\end{equation}
The imposed $\beta_1, \beta_2$ constraints
\eqref{eq:2:1:8}--\eqref{eq:2:1:10} have a well-interpretable
graphical visualization as shown in Fig. \ref{fig:21}. The set of
admissible $\beta_1, \beta_2$ values is inside of a color-shadowed
triangle. The $\beta_1 = \beta_2$ edge recovers the original
sub-optimal SM control. Obviously, the $D/U$ ratio determines the
size of the admissible $\{\beta_1, \beta_2\}$ set. It also
determines whether the negative $\beta_2$-values are allowed,
while this can lead to a twisting mode during the convergence, see
\cite{ruderman2023energy}. Recall that it is on the interval
$(\beta_2 \sigma_{M}, \, \beta_1 \sigma_{M})$ where the control
value is switched off, i.e. $u=0$, and thus operates as
energy-saving. It appears at each reaching cycle, i.e. between two
consecutive extremal values, i.e. $\sigma_{M_i}$ and
$\sigma_{M_{i+1}}$, and that throughout the entire convergence
period.

\begin{figure}[!h]
\centering
\includegraphics[width=0.98\columnwidth]{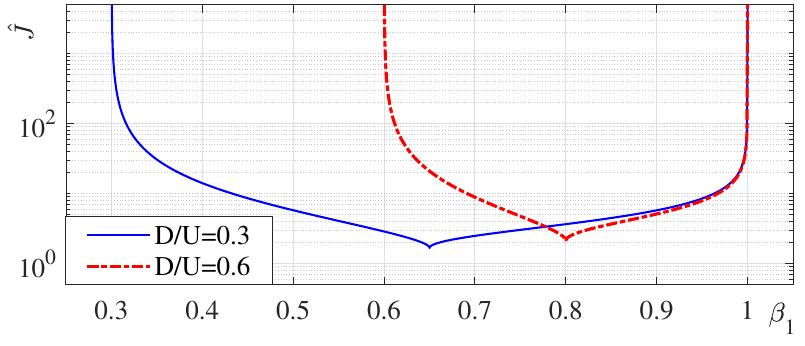}
\caption{Convergence time cost function $\hat{J}$ in dependency of
$\beta_1$.} \label{fig:22}
\end{figure}

\begin{figure}[!h]
\centering
\includegraphics[width=0.98\columnwidth]{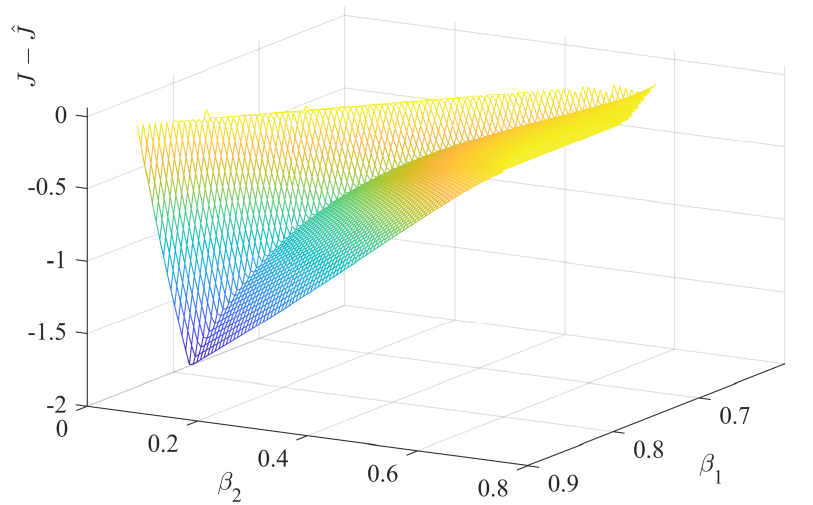}
\caption{Constrained objective function ($J$ -- $\hat{J}$) of the
$\beta_1$, $\beta_2$ parameters, for an exemplary
perturbation-to-control ratio $D/U=0.3$, ref.
\cite{ruderman2023energy}.} \label{fig:23}
\end{figure}

For an energy-saving operation of the SM control \eqref{eq:2:1:6},
a constrained optimization
\begin{equation}\label{eq:2:1:11}
\underset{\beta_1,\beta_2} {\min} \, \Bigl[  J(\beta_1, \beta_2,
U, D) - \hat{J}(\beta_1, U, D) \Bigr]
\end{equation}
was formulated and solved in \cite{ruderman2023energy}. This
maximizes the overall time difference of the control-on mode, i.e.
$u \neq 0$, between the original sub-optimal SM control, for which
the cost function is denoted by $\hat{J}$, and the energy-saving
SM control \eqref{eq:2:1:6}, for which the cost function is
denoted by $J$. Note that for both cost functions the $\beta_1$
parameter is kept fixed, also under the specified upper bound
\begin{equation}\label{eq:2:1:12}
\hat{J}(\beta_1, U, D) < \hat{J}_{\max}.
\end{equation}
An explanatory $\hat{J}$ mapping is exemplary shown in Fig.
\ref{fig:22}, while a boundary value $0 < \hat{J}_{\max} < \infty$
can be seen as application-specific and pre-determined for
assigning $\beta_1$.
Also we notice that both cost functions constitute a worst-case
multiplicative factor is the estimated finite convergence time,
cf. \cite{ruderman2023energy}. The minimization hard constraint is
\begin{equation}\label{eq:2:1:13}
J(\beta_1, \beta_2, U, D) - \hat{J}(\beta_1, U, D) < 0,
\end{equation}
while for each $D/U$ ratio, an optimal $(\beta_1, \beta_2)$ pair
can be calculated, see \cite{ruderman2023energy}. The results of
such constrained optimization \eqref{eq:2:1:11}--\eqref{eq:2:1:13}
are exemplary shown in Fig. \ref{fig:23} for the assumed $D/U=0.3$
ratio.

For a detailed analysis of the convergence and efficiency of the
energy-saving sub-optimal SM control, an interested reader is
referred to the full paper \cite{ruderman2023energy}.

\subsection{Second-order terminal SM control} \label{sec:2:sub:2}

For second-order systems, i.e. with the dynamic states $x_1(t)$
and $x_2(t) = \dot{x}_1(t)$, the so called terminal SM control
\begin{eqnarray}
\label{eq:2:2:1}
u(x_1, x_2) &=& -K_t \, \mathrm{sign}\bigl( \sigma(\cdot) \bigr), \\
\sigma(x_1, x_2) &=& x_1 + \delta \,  x_2^2 \, \mathrm{sign}
(x_2), \label{eq:2:2:2}
\end{eqnarray}
was proposed in \cite{zhihong1994}, while a quadratic sliding
surface $\sigma(\cdot)$ similar to \eqref{eq:2:2:2} was
exemplified already before, see \cite[Sec.~4]{levant1993}.
Remarkable in \eqref{eq:2:2:2} is that for an appropriately
selected $\delta$, the parabolic sliding surface coincides with
the time-optimal state trajectory of the well-known Fuller's
problem \cite{fuller1960relay}. This has been further analyzed for
the second-order motion systems in \cite{ruderman2021} so that for
\begin{equation}\label{eq:2:2:0}
\delta = \alpha  / U,
\end{equation}
where $U \equiv \max |u| $ is the maximal control amplitude, the
$\alpha = 0.5$ value represents the boundary layer of the terminal
mode, cf. \cite[Fig.~2]{ruderman2021}. That means for $0 < \alpha
< 0.5$, the control \eqref{eq:2:2:1}, \eqref{eq:2:2:2} changes to
the so-called \emph{twisting mode}, see e.g.
\cite{shtessel2014,utkin2020} for details. Both modes, the
terminal and twisting, provide a global finite-time convergence of
the $(x_1,x_2)$ trajectories to the origin. Also recall that for
the control \eqref{eq:2:2:1}, \eqref{eq:2:2:2} remains globally
finite-time convergent in case of the bounded perturbations
$|d(t)| < D$, the following control-gain condition $K_t > D$ must
be fulfilled.

With regard to the above basics, the used second-order terminal SM
control is given by
\begin{eqnarray}
\label{eq:2:2:3}
\bar{u} &=& -U \, \mathrm{sign} \, ( \sigma ), \\
\sigma &=& x_1 + U^{-1} \alpha \,  x_2^2 \, \mathrm{sign} (x_2),
\label{eq:2:2:4}
\end{eqnarray}
while $\alpha > 0.5$ must be assigned with respect to $D$ if an
avoidance of the twisting mode must be guaranteed. Also worth
noting is that if the second-order dynamics is additionally scaled
by an inertial term $m$, as in the experimental system provided in
Section \ref{sec:3}, then the sliding surface \eqref{eq:2:2:4}
must additionally include $m$, cf. with
\cite[eq.~(13)]{ruderman2021}.

The same quadratic terminal sliding surface \eqref{eq:2:2:4},
which provides a finite-time convergence to origin, can equally be
used in combination with other SM control laws, i.e. differing
from \eqref{eq:2:2:3}, as it is done with the energy-saving
sub-optimal SM control below during experimental evaluation.

\subsection{Chattering due to additional actuator dynamics} \label{sec:2:sub:3}

The so called \emph{chattering}, as a well-known phenomena in SM
control systems, see e.g. in \cite{shtessel2014,utkin2020},
appears due to unmodeled dynamics (often of the sensors and
actuators) and, thus, arises naturally when SM controllers are
implemented. Usually, it results in high-frequency oscillations
with a finite amplitude. Often, the not modeled (and rather
parasitic) additional dynamics can be captured by a first-order
behavior
\begin{equation}\label{eq:2:3:1}
v(s) = (\mu s + 1)^{-1} u(s),
\end{equation}
written here in Laplace domain with the complex variable $s$. The
corresponding time constant $\mu > 0$ can well approximate the
cumulative first-order time delay of both, the sensor and actuator
in the loop. Then, the control value $v(t)$ appears as the input
channel of the system instead of $u(t)$, cf. \eqref{eq:2:2}, thus
increasing relative degree of the sliding variable.

For the energy-saving sub-optimal SM control \eqref{eq:2:1:6}, the
analysis of chattering was addressed in \cite{ruderman2023energy},
following similar lines of development as provided in
\cite{boiko2007}. The solution of the harmonic balance equation,
which involves also the actuator dynamics \eqref{eq:2:3:1}, yields
the chattering frequency
\begin{equation}\label{eq:2:3:5}
\tilde{\omega} = \mu^{-1} \frac{\beta_1+\beta_2}{
\sqrt{1-\beta_1^2}+\sqrt{1-\beta_2^2}}.
\end{equation}
The corresponding amplitude of steady-state oscillations is
\begin{equation}\label{eq:2:3:6}
A_{\sigma} = \frac{\sqrt{\mu^2 \tilde{\omega}^2 +
1}}{\tilde{\omega}^2 \bigl( \mu^2 \tilde{\omega}^2 + 1 \bigr) }.
\end{equation}

For the terminal SM control, the appearance of chattering is as
inherent as for the conventional first-order SM control with
discontinuous control action, and has an even higher amplitude,
see \cite{utkin2015}. This appears natural since for the terminal
surface \eqref{eq:2:2:2}, when linearized in vicinity to origin,
the gaining factor of the linearized subsystem in the loop
increases largely. This leads, when using a harmonic balance
analysis, see e.g. \cite{boiko2009}, to an interception of the
corresponding Nyquist plot by the relay's describing function at
already lower frequencies comparing to the linear first-order
sliding surface. Therefore higher oscillation amplitudes appear. A
more detailed discussion of chattering in the terminal SM control
is not included in this work due to a different focus.

\section{Benchmarking physical system}
\label{sec:3}

\subsection{Voice-coil based linear actuator} \label{sec:3:sub:1}

\begin{figure}[!h]
\centering
\includegraphics[width=0.45\columnwidth]{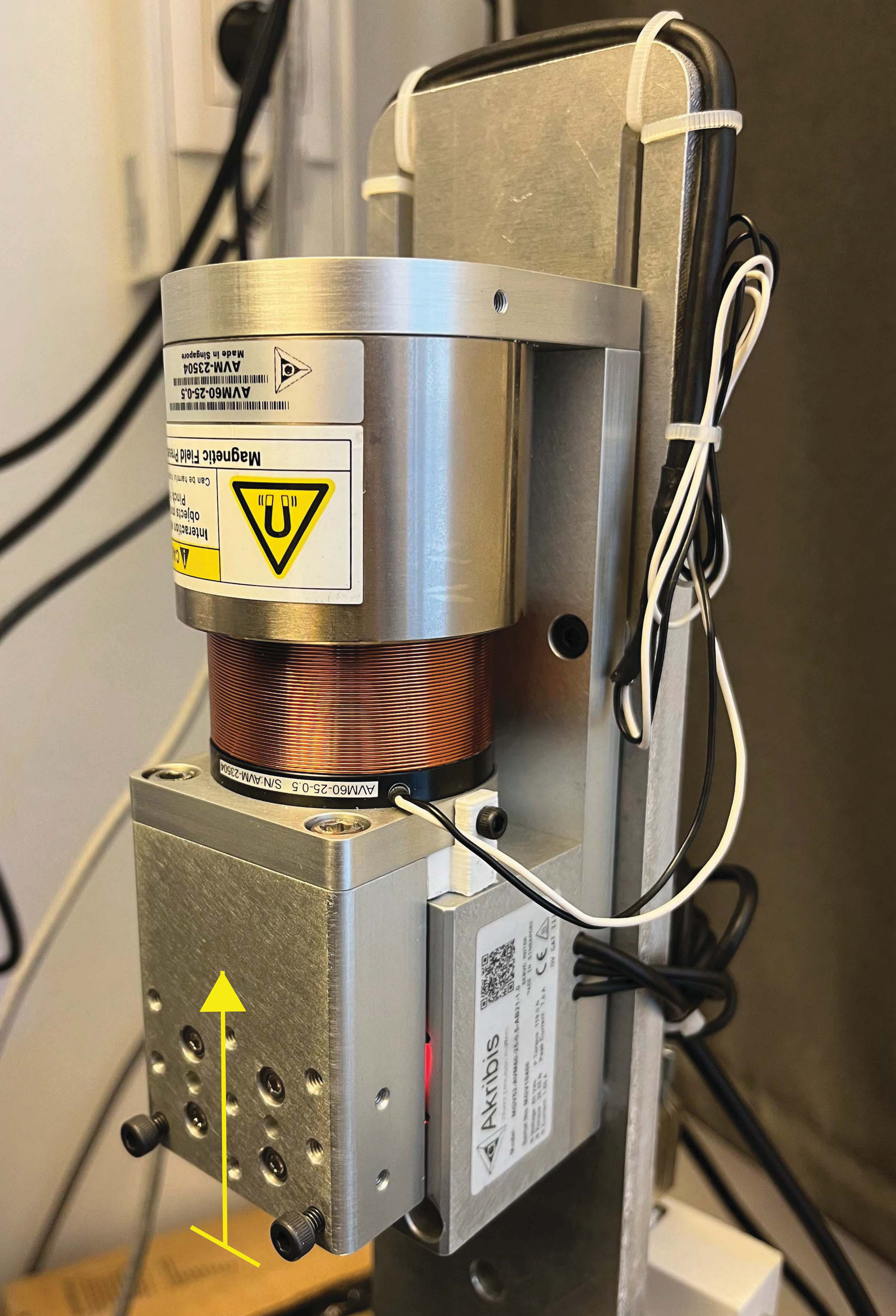}
\caption{Voice-coil based linear actuator with one DOF in $x$
coordinates.} \label{fig:expsetup}
\end{figure}

The benchmarking experimental system in use is a voice-coil based
actuator, see e.g. \cite{okyay2014} for basics, with one linear
degree of freedom (DOF). The laboratory view of the actuator is
shown in Fig. \ref{fig:expsetup}. More technical details,
including the system parameters, can be found in
\cite{ruderman2022motion}. Most relevant to be noted is that the
sampling rate $f_s$ of the real-time control is relatively high,
set to 10 kHz. And the time derivative of the measured output
displacement $x(t)$ is obtained by the discrete time
differentiation combined with a low-pass filter (LPF). The cutoff
frequency is 1 kHz. Recall that $\dot{x}(t)$ is required for
constructing the sliding surface $\sigma$, cf. \eqref{eq:2:2:4}.

The modeled system dynamics is given by
\begin{eqnarray} \label{eq:3:11}
  m \ddot{x}(t) &=& v(t) + G + d(t), \\
  \mu \dot{v}(t) &=& K u(t) - v(t).
\label{eq:3:12}
\end{eqnarray}
The overall mass of the actuator mover is $m=0.538$ kg. The input
gaining factor, which takes into account both the
electro-magneto-mechanical coupling constant and the resistance of
the anchor coils, is $K = 3.28$ N/V. The constant gravity term is
$G = 5.27$ N, while the cumulative unknown bounded disturbance is
$0 < \bigl|d(t)\bigr| < D$. Estimated from various open- and
closed-loop experiments, the upper bound of disturbances is set to
$D = 1$ N. The time constant of the electro-magnetic circuit of
the actuator $\mu = 0.0012$ sec is taken from the manufacturer's
technical data sheet. The available signals are the input voltage
$u(t)$ (given in V) and the output relative displacement $x(t)
\equiv x_1(t)$ (given in m).

\subsection{Noisy system states} \label{sec:3:sub:2}

\begin{figure}[!h]
\centering
\includegraphics[width=0.98\columnwidth]{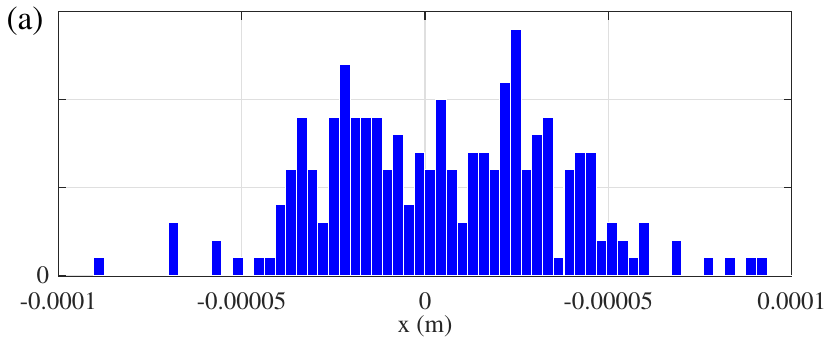}
\includegraphics[width=0.98\columnwidth]{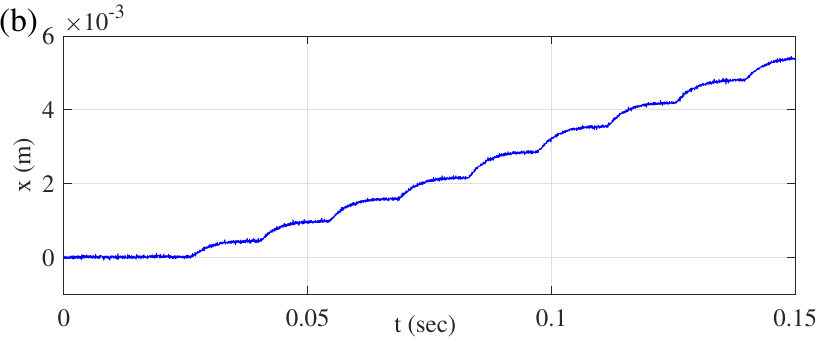}\\[3mm]
\includegraphics[width=0.98\columnwidth]{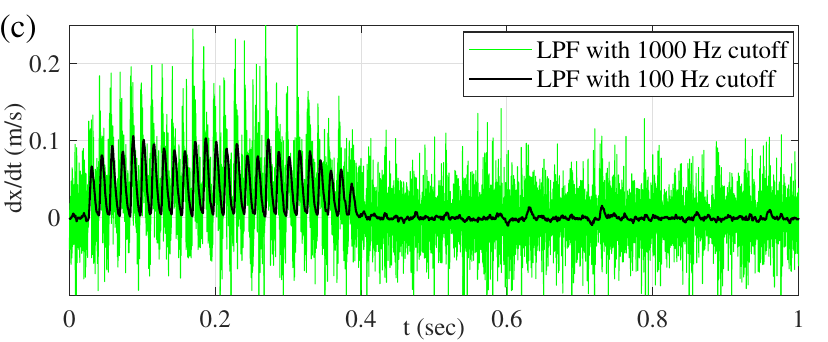}
\caption{Noisy system signals: distribution of the measured output
$x(t)$ at zero input $u(t) = 0$ in (a), fragment of the measured
output $x(t)$ at applied constant input $u(t)=\mathrm{const}$ in
(b), fragment of the low-pass filtered $dx/dt$ obtained by the
discrete time differentiation of the measured $x(t)$ in (c).}
\label{fig:noisysignal}
\end{figure}

The measured output signal is subject to the noise. This is due to
a contactless sensing of the relative displacement by means of the
inductive sensor, while its nominal repeatability is $\pm 12$
micrometer. When the control signal is off, i.e. $u(t) = 0$, the
background noise of the output is still visible. The distribution
of an exemplary measured output $x(t)$ at zero input is shown in
Fig. \ref{fig:noisysignal} (a). While zero mean-value can be
assumed, a nearly normal distribution can be recognized, and the
evaluated standard deviation is $4.58 \times 10^{-5}$ m.
To get an idea about additional process noise, i.e. disturbances,
one can inspect a fragment of the measured output $x(t)$ at the
applied constant input $u(t)=\mathrm{const}$, see Fig.
\ref{fig:noisysignal} (b). One can recognize a harmonic pattern in
the output relative displacement, caused by a electro-magnetic
cogging of the stator-armature structure. Worth noting is that the
disturbance $d(t)$ includes also the mechanical friction effects,
see e.g. \cite{ruderman2023analysis}, and the position-dependent
variations of the gaining factor $K$, see e.g. \cite{voss2022}.
The velocity state $\dot{x}(t)$, required for computing the
sliding variable $\sigma(t)$, is obtained by the discrete-time
derivative of the measured output. Subsequently, the second-order
LPF is applied for obtaining a smoothed velocity $w(t)$. The LPF
dynamic behavior is given by
\begin{equation}\label{eq:3:2}
\ddot{w}(t) + 2 \omega_{c} \, \dot{w}(t) + \omega_{c}^2 \, w(t) =
\omega_{c}^2 f_s \, \Bigl( x\bigl(t\bigr)-x\bigl( t- \frac{1}{f_s}
\bigr)\Bigr),
\end{equation}
where $\omega_{c} = 2 \pi f_c$ is the LPF cut-off frequency. The
low-pass filtered velocity state is exemplary shown in Fig.
\ref{fig:noisysignal} (c), and that for a controlled motion in the
time between 0.02 sec and 0.4 sec. Two signal patterns are shown
opposite to each other for the LPF with $f_c = \{100, \, 1000\}$
Hz. The velocity signal filtered with $f_c = 100$ Hz is depicted
for visualizing both, a continuous motion affected by the cogging
force disturbance and a noisy velocity in the idle state, i.e. for
$t > 0.4$ sec. On the contrary, the velocity signal filtered with
$f_c = 1000$ Hz is actually used in the control experiments when
calculating the sliding variable $\sigma(t)$. Here it is worth
recalling that the settling time $t_s$ of the step response of the
second-order LPF system \eqref{eq:3:2}, i.e. $|\dot{x}(t) - w(t)|
\cdot |\dot{x}(t)|^{-1} < 0.01$ for $t > t_s$, can be
approximately calculated by
$$
t_s \approx 4.6 \, \omega_c^{-1}.
$$
One can recognize that the settling time of the LPF with 100 Hz
cut-off frequency is $t_{s,100} = 0.0073$ sec and, thus, about 6
times larger than the actuator time constant $\mu$. On the
contrary, the settling time of the LPF with 1000 Hz cut-off
frequency is $t_{s,1000} = 0.00073$ sec and, thus, lies clearly
below the actuator time constant $\mu$. Therefore, it appears as
sufficiently low for use of $w(t) \approx dx/dt$ when calculating
the sliding variable $\sigma(t)$.

\section{Experimental results}
\label{sec:4}

The experimental results demonstrated below are obtained by using
the control value
\begin{equation}\label{eq:4:1}
u(t) = u_{sm}\bigl(x,\dot{x}\bigr) + u_{gr},
\end{equation}
where $u_{gr} = 1.61$ V is compensating for the constant gravity
term, cf. \eqref{eq:3:11}, and the sliding-mode control $u_{sm} =
\{\bar{u}, \, \tilde{u}\}$ is either the terminal \eqref{eq:2:2:3}
or the energy-saving sub-optimal one \eqref{eq:2:1:6}, both using
the same second-order sliding surface \eqref{eq:2:2:4}. The
magnitude of the switching control is assigned to $U = 0.8$ V,
that is about 2.5 times higher than the corresponding upper bound
of disturbances, cf. with $D$ in section \ref{sec:3:sub:1} and
\cite{ruderman2023energy}. The design parameter of the terminal
sliding surface is assigned as $\alpha = 1.2$, thus preserving the
terminal algorithm from the twisting mode with respect to the
perturbations upper bound $D$. The threshold parameters of the
energy-saving sub-optimal control are assigned to be $\beta_1 =
0.85$ and $\beta_2 = 0.1$, respectively, cf.
\cite{ruderman2023energy}.

The control experiments were conducted for the reference step $x_r
= 0.015$ m, starting from the same zero relative position. Note
that through a transformation of coordinates $x^* = x - x_r$ the
step reference control task becomes equivalent to stabilization in
the origin $(x^*, \dot{x})=(0,0)$. Below, we will however continue
to use $x$ instead of $x^*$, this for the sake of consistency.

\begin{figure}[!h]
\centering
\includegraphics[width=0.99\columnwidth]{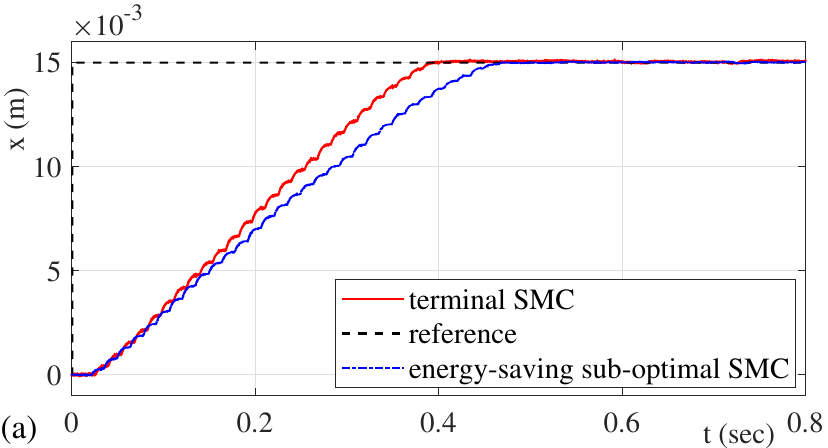}
\includegraphics[width=0.99\columnwidth]{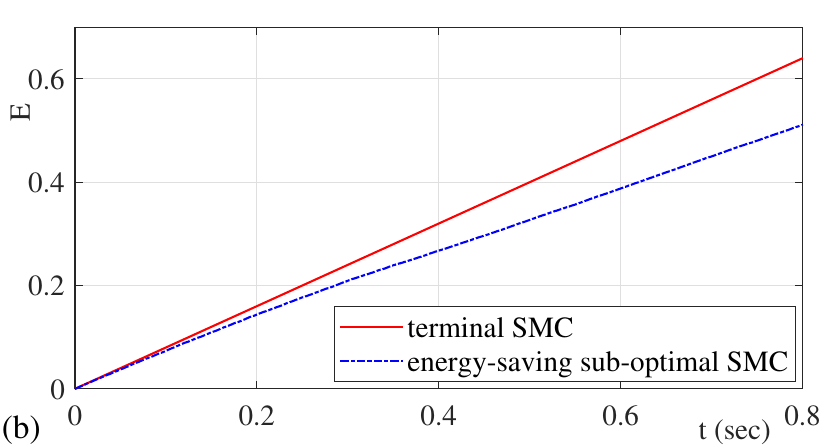}
\caption{Measured output response of the SM control $\bar{u}$ and
$\tilde{u}$ in (a), and the corresponding integral $E$ of the
control absolute control signal in (b).} \label{fig:expresult}
\end{figure}

For evaluation of the control energy consumption, the following
integral metric is used
\begin{equation}\label{eq:4:2}
E = \int |u(t)| dt.
\end{equation}
Recall that for a continuously switching SM control $|u| = U$,
while for the energy-saving sub-optimal SM control $|u| = \{U, \,
0 \}$. Therefore, the metric \eqref{eq:4:2} reflects the overall
time when the control signal is on and, hence, the energy
consumption during the controlled operation. The measured output
response of both SM controllers under benchmarking is shown Fig.
\ref{fig:expresult} (a).
Both controls reach the reference equilibrium in finite time and
show the same steady-state accuracy, while the reaching time of
the terminal SM control is slightly superior comparing to the
energy-saving sub-optimal SM control, cf.
\cite{ruderman2023energy}. This is not surprising since both
controllers maintain the system on the same sliding manifold
\eqref{eq:2:2:4}, while the terminal SM controller is always on.
The benefit of the energy-saving sub-optimal SM control becomes
visible when inspecting the evaluated metric \eqref{eq:4:2}
depicted in Fig. \ref{fig:expresult} (b). Not only during the
reaching phase, i.e. on the time interval bounded by approximately
$t < 0.4$ sec, but equally after the convergence the difference in
energy consumption between both SM controllers continue to grow.

Another interesting fact is also related to the sliding-mode
chattering, which can appear due to an additional (parasitic)
actuator dynamics, cf. section \ref{sec:2:sub:3}. Evaluating the
chattering frequency and amplitude by means of \eqref{eq:2:3:5},
\eqref{eq:2:3:6} results in
$$
\tilde{\omega} = 520.2 \hbox{ rad/sec}, \quad   A_x \approx 3.1
\times 10^{-6} \hbox{ m}.
$$
Note that $A_x$ can easily been estimated from $A_{\sigma}$, cf.
\eqref{eq:2:3:6}, in vicinity to origin. Given the sensor
resolution and level of the measurement noise, cf. section
\ref{sec:3}, the above evaluated chattering parameters do not
allow to detect a harmonic oscillations pattern within the
measured output $x(t)$. Similar conclusions can be drawn and
observed experimentally for the second-order terminal SM control
in regard to the used actuator with additional (parasitic)
dynamics.

\section{Conclusions} \label{sec:5}

The energy-saving sub-optimal SM control, recently proposed in
\cite{ruderman2023energy}, was for the first time evaluated
experimentally and benchmarked with a conventional discontinuous
SM control, both operating on the same terminal sliding surface.
The surface ensures both SM controllers to operate in the
second-order sliding mode, while the latter appears in a finite
time. The real experimental system with the measurement and
process noise, and without sensing of the time derivative of the
output and of the sliding variable, is considered. The system is
modeled in a simplified way by a perturbed double integrator. The
latter is with the known upper bound of the matched perturbations
and an additional (parasitic) first-order actuator dynamics. The
same switching control magnitude and the designed terminal sliding
surface allow for a fair comparison of both, the conventional
discontinuous (terminal) and energy-saving sub-optimal SM
controllers. The integral of the absolute control value is
evaluated as an adequate energy consumption metric which is
proportional to the overall time when the control with a constant
magnitude was on. The provided experimental case study confirms
the applicability and efficiency of the energy-saving sub-optimal
SM control \cite{ruderman2023energy}.

\bibliographystyle{IEEEtran}
\bibliography{references}

\end{document}